\documentclass[a4paper,11pt,oneside,fleqn]{article}

\usepackage[mathscr]{eucal}
\usepackage{amsmath,amssymb,amsthm}
\usepackage{graphicx}
\usepackage{subcaption}
\usepackage{url}

\usepackage{cite}
\usepackage{hyperref}
\hypersetup{colorlinks, linkcolor=blue, citecolor=blue, urlcolor=blue}

\allowdisplaybreaks
\makeatletter
\long\def\@makecaption#1#2{%
  \vskip\abovecaptionskip\footnotesize
  \sbox\@tempboxa{#1. #2}%
  \ifdim \wd\@tempboxa >\hsize
    #1. #2\par
  \else
    \global \@minipagefalse
    \hb@xt@\hsize{\hfil\box\@tempboxa\hfil}%
  \fi
  \vskip\belowcaptionskip}
\makeatother

\newcommand{\todo}[1][\null]{\ensuremath{\clubsuit}}

\newcommand{\noprint}[1]{}
\newcommand{\checked}[1][\null]{\ensuremath{\boldsymbol{\surd}}}

\newtheorem{theorem}{Theorem}

\newtheorem*{problem*}{Problem}
{\theoremstyle{definition}

\newtheorem{remark}[theorem]{Remark}
\newtheorem*{remark*}{Remark}
}

\begin{document}

\par\noindent {\LARGE\bf
A generative adversarial network approach\\
to (ensemble) weather prediction\par}

\vspace{5mm}\par\noindent{\large
Alex Bihlo$^\dag$
}

\vspace{5mm}\par\noindent{\it
$^\dag$\,Department of Mathematics and Statistics, Memorial University of Newfoundland,\\
$\phantom{^\dag}$\,St.\ John's (NL) A1C 5S7, Canada
}

\vspace{6mm}\par\noindent
E-mail:
abihlo@mun.ca

\vspace{7mm}\par\noindent\hspace*{8mm}\parbox{140mm}{\small
We use a conditional deep convolutional generative adversarial network to predict the geopotential height of the 500 hPa pressure level, the two-meter temperature and the total precipitation for the next 24 hours over Europe. The proposed models are trained on 4 years of ERA5 reanalysis data from 2015--2018 with the goal to predict the associated meteorological fields in 2019. The forecasts show a good qualitative and quantitative agreement with the true reanalysis data for the geopotential height and two-meter temperature, while failing for total precipitation, thus indicating that weather forecasts based on data alone may be possible for specific meteorological parameters. We further use Monte-Carlo dropout to develop an ensemble weather prediction system based purely on deep learning strategies, which is computationally cheap and further improves the skill of the forecasting model, by allowing to quantify the uncertainty in the current weather forecast as learned by the model.
}\par\vspace{5mm}


\section{Introduction}

As almost all fields of sciences, also meteorology has seen a steady increase in interest in using deep learning over the past few years. Traditional weather prediction is done with advanced numerical models that numerically solve the partial differential equations believed to govern the atmosphere--ocean system. This is a considerably costly endeavour, with improvements in forecasting quality typically critically linked to the improvements in available computing resources. Besides being computationally expensive, traditional numerical weather forecasts also require advanced physical parameterization schemes for unresolved processes, see e.g.~\cite{sten07Ay}, that have to be tailored each time the model resolution is increased. Weather forecasting is also a data intensive endeavour. Various instruments (satellites, air planes, radiosondes, synoptical stations) continuously measure the state of the atmosphere, and have led to an excessive amount of data available for the past 30 years and more. For more information and background on the various models that have been used in numerical weather prediction, see e.g.~\cite{holt04a}.

While the steady progress in obtaining more capable supercomputers, improving the network of observational systems, and strides being made in improving physical parameterization schemes have rendered numerical weather prediction a resounding success story, see e.g.~\cite{baue15a}, the prospect of using machine learning to interpret weather prediction as a supervised learning problem has started to receive considerable attention in the meteorological community, see e.g.~\cite{sche19a, weyn19a} for some initial results. While costly to train, once production ready deep neural networks would conceivably allow issuing numerical weather predictions at a fraction of the computational cost of traditional differential equations based forecasts. While still in its infancy, and presently no deep learning based model being able to consistently beat conventional numerical weather prediction models, first results have been encouraging and are preparing the grounds for further research in this area. In the present paper we are interested in two research problems. 

We firstly aim to investigate the ability of a conditional generative adversarial (cGAN) network based architecture to learn the basic conditional distribution underlying the atmospheric system. In other words, we aim to investigate how well cGANs can learn the physics underlying meteorological data. GANs have shown substantial capabilities in a variety of image-to-image mapping problems, see e.g.~\cite{isol17a}. Interpreting weather prediction as a video-to-video mapping problem provides a direct framework accessible to GANs. The underlying data we train our network against are the state-of-the-art ERA5 reanalysis data, which are produced by the European Centre for Medium Range Weather Forecasts (ECWMF), and available from 1979 to within 5 days of real time, see~\cite{c3s17a}.

The second problem we are interested in is the problem of quantifying the uncertainty in weather forecasts using machine learning. Successful weather predictions have to deal with uncertainties, which enter at various levels: Uncertainties in the initial state of the atmosphere (the initial conditions), uncertainties in the model formulation itself, and uncertainties inherent in the atmosphere itself, see~\cite{bern17a,lore63Ay,lore96a}. The common approach to tackle this problem of uncertainty in weather forecasts is ensemble prediction, where carefully thought out modifications are introduced in the initial data and the model itself. Running multiple instances of a forecasting model for a variety of initial conditions allows one to produce an ensemble of forecasts, which provides statistical information about the state of the atmosphere. Depending on the predictability of the current state of the atmosphere, this ensemble will have smaller or larger spread, with smaller (larger) spread ideally signifying higher (lower) predictability. Here we aim to investigate whether meaningful statistical information could be obtained by running an ensemble of slightly different cGANs. One computationally cheap and straightforward way of accomplishing this is to introduce dropout layers into the model and use dropout both at the training and testing stage, i.e.\ by using Monte-Carlo dropout, see~\cite{gal16b}. In other words, we will investigate whether the statistical information produced using Monte-Carlo dropout aligns with meaningful physical uncertainty of the future state of the atmosphere.

As features we aim our cGAN models to learn we choose the \textit{geopotential height of the 500 hPa pressure layer}, the \textit{two-meter temperature} and the \textit{total precipitation}. We use the same model architecture to separately predict these features. That is, we train one cGAN each for each feature, which uses as training data that feature alone (e.g.\ the precipitation cGAN is trained using precipitation data alone). The choice for these parameters is motivated by the following reasoning. 

The height of the 500 hPa pressure layer is among the easiest meteorological quantities to predict. Some 70 years ago, Jule Charney, Ragnar Fj{\o}rtoft and John von Neuman succeeded in using the ENIAC computer for producing the first successful numerical weather prediction, see~\cite{char50a}\footnote{For a historical account as well as a re-creation of the original ENIAC integrations, see~\cite{lync08a}.}. Due to the computational limitations at the time, they decided to predict the 500 hPa geopotential height using the so-called barotropic vorticity equation. The advantage of this equation is that it is a single partial differential equations predicting the height of the two-dimensional isosurface of the 500 hPa geopotential layer, thus alleviating the computational burden that would have incurred had the fully three-dimensional set of governing equations been used. In addition, this equation filters out the fast moving gravity waves that would have required at that time impossibly short time steps of the numerical integrations, while preserving the much slower moving Rossby waves, that largely govern the large scale weather patterns in the midlatitudes. Indeed, it is surprising how well the simple barotropic vorticity equation can capture the evolution of the 500 hPa pressure level. From a machine learning perspective this is ideal in that it gives hope that a model trained on 500 hPa geopotential height data alone may be able to successfully predict the future evolution of the 500 hPa geopotential height. As such, and following the works of~\cite{sche19a, weyn19a}, we regard it as an excellent benchmark for data-driven weather forecasting. 

The two-meter temperature is of obvious practical interest when it comes to every-day weather predictions. In contrast to the geopotential height of the 500 hPa pressure layer, which at least can be approximately predicted in a stand-alone fashion, the two-meter temperature is the product of the interaction of a multitude of physical processes. Radiation (both incoming short-wave and outgoing long-wave), clouds, atmospheric boundary-layer effects, topography, and proximity to large bodies of water amongst others have a substantial impact on the local distribution of two-meter temperature, see~\cite{sten07Ay}. While at the available resolution of the ERA5 data many of these processes are not adequately captured, we still expect to be able to assess the ability of the model to learn the impact of topographic effects as well as land--sea discrepancies on the forecast. 

The total precipitation is chosen as it is traditionally challenging to predict, even with state-of-the-art operational numerical weather models. In particular, precipitation is a by-product of processes such as atmospheric convection which is notoriously hard to represent in weather prediction models, see e.g.~\cite{yano18a}. As such, we do not expect a cGAN trained using precipitation alone to be successful in accurately predicting precipitation, no less since the ERA5 reanalysis data used are not high resolution enough to properly resolve the spatial distribution of precipitation. Regardless, we choose this parameter to get an initial idea about the limitations of data-only driven weather forecasts.

The further organization of this paper is the following. In Section~\ref{sec:PreviousWork} we recount different approaches that have been used in weather prediction over the past several decades and frame the machine learning based approach to weather forecasting as a video-to-video mapping problem. Section~\ref{sec:Model} contains a description of the proposed cGAN models for one-shot, multi-step weather forecasting, along with a description of the data being used to train these models. In Section~\ref{sec:Results} we present the results of this data-driven investigation, which includes both a computation of standard meteorological forecast quality scores as well as several case studies that provide some insights into whether or not the ensemble cGAN models have managed to learn some of the underlying physics of the atmosphere and its inherent uncertainty. The final Section~\ref{sec:Conclusions} contains the conclusions of this paper along with thoughts on possible future research directions in this field.

\section{Previous work}\label{sec:PreviousWork}

In contrast to data-only driven forecasting approaches that are commonly used in machine learning, in the mathematical sciences the forecasting problem is typically tackled by solving differential equations. In meteorology, this has led to a cascade of increasingly complicated models, beginning with the above mentioned barotropic vorticity equation, to quasi-geostrophic multi-layer models and the incompressible primitive equations, to the fully compressible three-dimensional governing equations of hydro-thermodynamics, see~\cite{holt04a}. The latter are the standard model for modern operational weather forecasting models.

Purely data-driven approaches to weather prediction have been investigated in~\cite{sche19a} to predict the 500 hPa geopotential height and 800 hPa temperature using a standard two-dimensional convolutional neural network. Similarly, in~\cite{weyn19a}, various flavours of two-dimensional convolutional neural network in combination with LSTMs were trained to predict the 500 hPa geopotential height over the northern hemisphere. More recently, in~\cite{weyn20a} the authors have extended their results to the whole Earth using convolutions on a cubed sphere with a U-net type convolutional neural network. Data-driven approaches also play an important role for weather nowcasting, i.e.\ weather forecasts with very short lead times, for which traditional numerical models can struggle to produce results of the high accuracy needed for decision makers such as airport planing authorities. See e.g.~\cite{bihl19a,xing15a} for the use of convolutional LSTM models for precipitation nowcasting. Machine and deep learning methods have also been used to post-process deterministic weather forecasts, see e.g.~\cite{rasp18a}, to down-scale climate data, see e.g.~\cite{moua17a}, and to improve sub-grid scale parameterization schemes, see e.g.~\cite{gent18a}. Related to the present study, the actual problem of ensemble weather prediction was studied by~\cite{sche18a}, where deep neural networks were trained on ensemble weather forecasts produced by an operational global ensemble forecasting system to learn the uncertainty in those forecasts. Recently~\cite{sche20a} have also investigated the problem of ensemble weather prediction by re-training a deep neural network multiple times with randomly initialized model weights yielding an ensemble of such models.

Sitting in between purely data-driven approaches and differential equation based models are physics-informed neural network approaches, see~\cite{rais19a}. Here one trains a deep artificial neural network to match both the given data and the data-generating underlying differential equation. This approach can also be used to identify a possibly hidden differential equation underlying the given data, see~\cite{rais18a}. In light of the success of differential equation based numerical weather prediction, we expect that pursuing a joint data and differential equation approach to weather prediction would be most sensible, and should be investigated in the future. 

However, here we restrict ourselves to a purely data-driven approach as is customary in machine learning for the following reasons: 
\begin{enumerate}\itemsep=0ex
\item Since we are using ERA5 reanalysis data that are produced from a version of the ECMWF operational weather prediction model, the underlying physical model that would have to be used for physics-informed neural networks would be extremely sophisticated, consisting of a complicated, fully three-dimensional system of nonlinear partial differential equations. In addition, it would be impossible to use the reanalysis data of the 500 hPa geopotential height, two-meter temperature and precipitation alone; all the variables governing the evolution of the atmospheric system would have to be used, including wind, temperature, pressure, and specific humidity at all atmospheric levels from the surface to the top of the atmosphere. This would be a quite substantial computational and technical challenge.

\item Since purely data-driven weather prediction has neither been investigated using a cGAN network nor using Monte-Carlo dropout for an ensemble prediction strategy, it is interesting to first study how accurate predictions this data-only approach can give. This will provide a first assessment of the capabilities of generative adversarial networks with dropout to learn the probability distribution of the physics and of the uncertainty underlying atmospheric dynamics.
\end{enumerate}

From a machine learning perspective we interpret the weather forecasting problem as a video prediction problem. More precisely, the past (observed) meteorological fields act as input frames and the goal is to predict the future meteorological fields as output frames. In other words, the weather forecasting problem is interpreted as a video-to-video mapping problem. Therefore, we term the models used in the sequel \texttt{vid2vid} models. 

Several architectures have been proposed to tackle this problem in the past, including variational autoencoder and generative adversarial networks, see for example~\cite{baba17a,dent18a,math15a,vond16a} for some results in this field. Two standing problems for video prediction are the separation of the static background from the dynamic foreground and the tendency of many networks to lead to blurry predictions at future frames. Luckily both of these issues are not critical for the weather forecasting problem. There is no static background in meteorological fields (in that the entire field is developing from each frame to the next), and since large-scale meteorological fields do not exhibit sharp gradients (or 'edges'), reasonably small amounts of blurring could be easily tolerated.

\section{Model architecture and training}\label{sec:Model}

In this section we provide the specification of the training and test data used, and provide details on the architecture of the cGAN models.

\subsection{Data}

\begin{figure}[!ht]
\centering
\begin{subfigure}{0.45\textwidth}
  \centering
  \includegraphics[width=\linewidth]{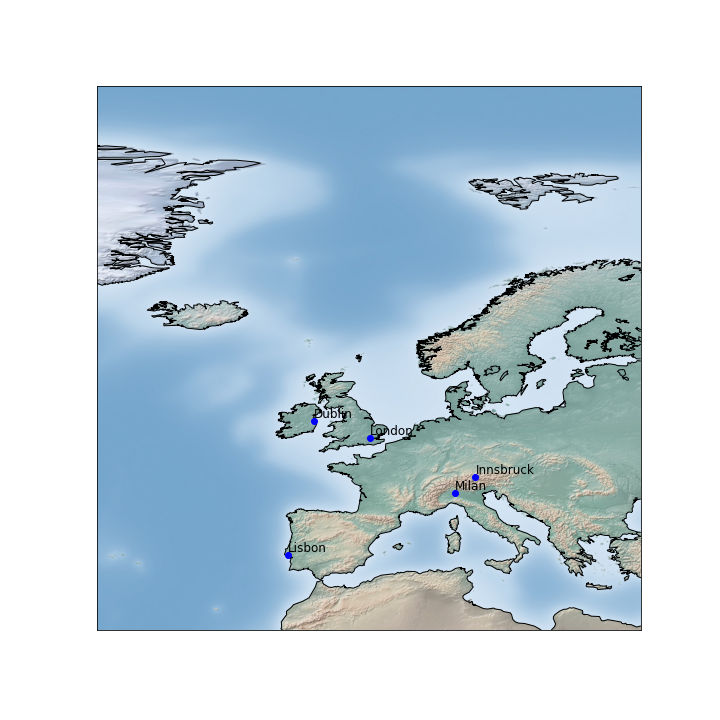}
  \caption{The domain in cylindrical projection.}
\end{subfigure}
\begin{subfigure}{0.45\textwidth}
  \centering
  \includegraphics[width=\linewidth]{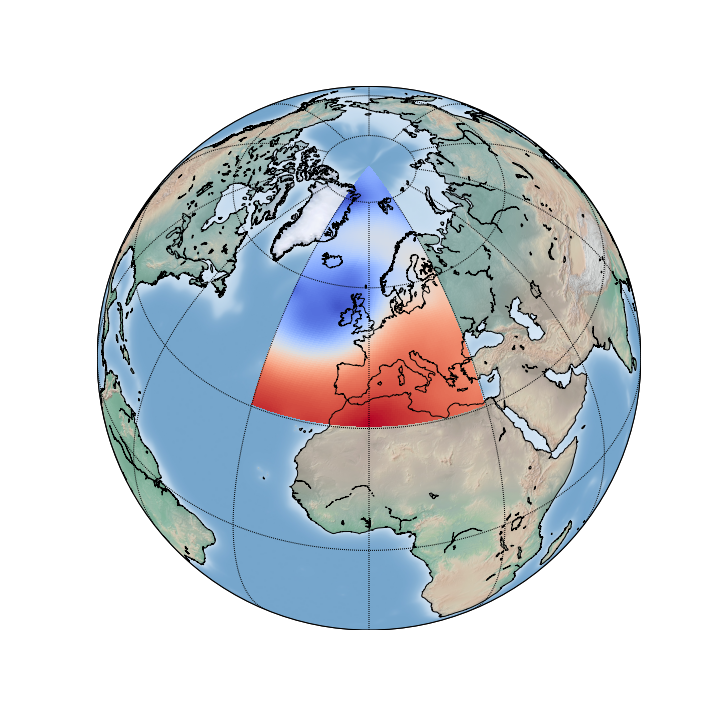}
  \caption{The domain in orthographic projection.}
\end{subfigure}
\caption{Outline of the computational domain.}
\label{fig:Domain}
\end{figure}

The input data against which our models were trained is the European Centre for Medium Range Weather Forecasts (ECMWF) Reanalysis version 5 (ERA5), see~\cite{c3s17a}. While this data is available for the entire Earth, we select a square in latitude--longitude coordinates extending from 30$^\circ$ North to 90$^\circ$ North and $-30^\circ$ West to 30$^\circ$ East as spatial domain, which covers most of Europe. See Figure~\ref{fig:Domain} for a depiction of the spatial domain both in cylindrical and orthographic projections. For the ease in representation, we choose the cylindrical projection in the following when presenting our results.

The original data is available on $0.25^\circ\times0.25^\circ$ resolution which for our domain yields a total of $240\times240$ spatial grid points. To speed up training we have down-sampled the data to $128\times128$ grid points, which thus corresponds to a resolution of about $0.5^\circ\times0.5^\circ$. For better reference, at 40$^\circ$N latitude, this corresponds to a spatial resolution of about $44$ km (longitude) times $56$ km (latitude). 

The native temporal resolution of the ERA5 data is hourly, but was chosen here to be 3 hours. Again we found that this provides a reasonable balance between the amount of data needed for training and typical time-scales on which notable frame-to-frame changes in large scale meteorological fields take place. We chose the period from 2015--2018 (4 years) for training and 2019 for testing.

All data is used in the original units as provided in the ERA5 datasets, which are $\mathrm{m}^2\mathrm{s}^{-2}$ for the geopotential height, Kelvin for temperature and meters for the total precipitation. Before training the cGAN models, we normalized the training data by subtracting the mean of the respective training data and dividing by the associated standard deviation. We reverse this normalization for plotting the results. For plotting and verification purposes, unless specified otherwise, we also change the units of these meteorological quantities to the more conventional meters ([m]) for the geopotential height (by dividing by the gravitational acceleration $g=9.81\, \mathrm{ms}^{-2}$), degree Celsius ([$^\circ$C]) for the temperature (by subtracting 273.15) and millimeters ([mm]) for the total precipitation (by multiplying by 1000).

\begin{remark}
Since we train our neural network against data corresponding to only a section of the Earth, the resulting model, in meteorological terms, corresponds to a so-called \textit{limited area model}. In contrast to global weather prediction models, a standing challenge for limited area models is the presence of boundary conditions. In a differential equations based limited area model, the flow of the prognostic quantities over the boundaries of the domain has to be prescribed over the forecast horizon. This is usually done by using the (typically lower-resolution), downscaled global model output at the boundaries over the forecast horizon as approximations for the boundary conditions, see~\cite{gior90a}. It is interesting to note that the present model does not require that such boundary conditions are provided. Due to training on large amounts of reanalysis data it can be expected that the model will learn the basic possible patterns of the 500 hPa geopotential height surface and two-meter temperature, and thus can reasonably well infer these boundary conditions for at least several hours of the prediction. Ideally, the model will also learn that the inflow boundary condition corresponds to an area of higher uncertainty, since the forecasting problem on a limited area domain without prescribed boundary conditions is ill-posed (in that there is no unique solution).
\end{remark}

\subsection{The model}

The cGAN model trained for this study is essentially a three-dimensional adaptation of the classical \texttt{pix2pix} architecture proposed in~\cite{isol17a}. The generator $G$ is a U-net architecture as first proposed by~\cite{ronn15a}, feeding in a sequence of \texttt{frames} past two-dimensional reanalysis data and aims to predict the next \texttt{frames} future two-dimensional reanalysis data. That is, rather than doing the forecast one step at a time, we predict the entire sequence of images at once. Here we choose \texttt{frames}=8, yielding 24 hours of input data $x$ to predict the next 24 hours of output data $y$. Therefore, input data $x$ and output data $y$ have dimensions $8\times 128 \times 128$ (time, longitude, latitude). We choose this one-shot forecasting approach over the iterative approach commonly used in video frame prediction, since weather predictions are typically issued at constant fixed time intervals for a constant forecasting window. 

Downsampling of the data in the U-net generator $G$ is done with three-dimensional strided convolutions with kernel sizes of $(4,4,4)$ and strides of either $(1,2,2)$ or $(2,2,2)$. All but the first layer use batch normalization. The activation chosen in each downsampling layer is leaky ReLU with $\alpha=0.2$. A total of six downsampling layers with 64, 128, 256, 256, 256, 256 filters, respectively, are used. The bottleneck of the generator model consists of two layers of convolutional LSTMs with kernel size $(2,2)$ and 256 filters each. Both dropout and recurrent dropout, see~\cite{gal16a}, are used for the LSTM bottleneck layers.

Upsampling is done using transposed three-dimensional strided convolutions with the analogous kernel sizes, strides and numbers of filter as used at the respective downsampling stage. ReLU is chosen as activation function in the upsampling blocks. Batch normalization is used as is dropout, except at the three last layers that do not use dropout. The last layer uses a linear activation function.

The discriminator $D$ is a PatchGAN similar as described by~\cite{isol17a} using the same encoder blocks as are used in the generator. A total of 5 layers with 64, 128, 256, 512 and 1 filters, respectively are used here. The last layer uses a sigmoid activation function.  

The associated GAN network aims to find the optimal generator $G^*$,
\begin{align*}
 &G^* = \mathrm{arg}\,\mathrm{min}_G\,\mathrm{max}_D\,\mathcal{L}_{\rm cGAN}(G,D) + \lambda \mathcal{L}_{\rm L1}(G),
\end{align*}
where
\begin{align*}
 &\mathcal{L}_{\rm cGAN}(G,D) = \mathbb{E}_{x,y}[\log D(x,y)] + \mathbb{E}_{x,s}[\log(1-D(x,G(x,s)))],\\
 & \mathcal{L}_{\rm L1}(G) = \mathbb{E}_{x,y,s}[||y - G(x,s)||_1],
\end{align*}
during training, i.e.\ we use L1 regularization of the network which is known to discourage blurring over the more standard L2 regularization. Here $s$ is random noise which is introduced by using dropout at both training \textit{and} test time, see~\cite{gal16b}. We have experimented with various dropout rates and found that too high a dropout inhibits the model to reasonably learn the inherent data distribution. A dropout rate of 20\% was experimentally found to give meaningful results and was thus used in all results reported below.

The model is trained using the \texttt{Adam} optimizer, see~\cite{king14a}, with a learning rate of 0.0002 and momentum parameters $\beta_1 = 0.5$ and $\beta_2 = 0.999$, respectively, and a batch size of 16. The network is trained if it has learned a mapping $G\colon (x,s)\to y$ producing results which to the discriminator $D$ will be indistinguishable from the real data for $y$. See~\cite{isol17a} for further details of the related \texttt{pix2pix} architecture and its training.

\begin{remark}
We should like to note here that 24 hours is a rather short forecasting time frame, in particular for classical ensemble predictions which are typically aimed at estimating the medium-range uncertainty in the atmospheric evolution. While the proposed \texttt{vid2vid} models could be easily trained for longer lead times as well, we refrain from doing so in the present study due to the different inherent scales in the meteorological quantities considered. While the geopotential height can be typically predicted accurately for several days in advance, see~\cite{baue15a}, the scales of typical precipitation events are much shorter, on order of a few hours to one day. To provide a unifying time frame for evaluating the various \texttt{vid2vid} models we choose a 24 hours forecast lead time as a balance between the longer scale geopotential height and the shorter scale precipitation evolutions. Longer forecast times would also require an explicit handling of the inflow boundary condition, which as discussed above is not present in the considered limited area \texttt{vid2vid} models.
\end{remark}

\subsection{Ensemble prediction using Monte-Carlo dropout}

It has been observed in~\cite{isol17a} that adding noise explicitly to the \texttt{pix2pix} architecture, as is done in standard unconditional GANs as proposed by~\cite{good14a}, was not an effective strategy since the model ignored the noise in its predictions. Rather, noise was added through the use of dropout at both training and test time, where it was observed that the resulting model behaved only minimally stochastically. While a highly stochastic output may be relevant for application ranges such as image-to-image translations (giving a variety of realistic images not already included in the training data), it is a fundamental question whether such wide stochasticity would be required for weather prediction. The degree to which the atmospheric system is stochastic and thus predictable is still up to debate, see~\cite{lore96a,zhan19a}, and most numerical weather prediction models are at their core deterministic, with stochasticity entering at the level of parameterizations and ensemble prediction techniques, see~\cite{baue15a,bern17a}. It is thus interesting to assess whether the modified cGAN architecture based on the work by~\cite{isol17a} can be effective in capturing the essential conditional distribution of the atmospheric parameters it models and their degree of inherent uncertainty.

We run an ensemble of 100 realizations of the respective \texttt{vid2vid} model for the chosen atmospheric parameters $z\in\{\rm{geo}, \rm{t2m}, \rm{tp}\}$ (corresponding to the fields of 500 hPa geopotential height, two-meter temperature and total precipitation, respectively). From this ensemble of results we then compute both the point-wise ensemble mean $z_{\rm mean}$ and standard deviations $z_{\rm std}$, respectively. Ideally, $z_{\rm std}$ should be a measure of the inherent uncertainty of the future state of the atmosphere, in that higher values of $z_{\rm std}$ in regions of the spatial domain should correspond to regions of higher uncertainty of the further atmospheric developments in that region.

Below we also show time series of the forecast parameters along with their associated probability cones at particular geographic locations. For determining the extend of the probability cones (the ensemble spread), i.e.\ the length of the interval $(z_{\rm mean}-\gamma z_{\rm std}, z_{\rm mean}+\gamma z_{\rm std})$ at each forecast time $t$, we need to specify the parameter $\gamma$. While ideally the standard deviation produced by the \texttt{vid2vid} model is directly correlated with the uncertainty in evolution of the current weather pattern, there is no inherent reason to believe that the magnitudes of these quantities would be the same.

Critically, the ensemble prediction model considered here only varies the model itself \textit{not} the initial data, which are always the same respective ERA5 reanalysis fields. Varying the initial data is classically a main ingredient in successfully capturing the correct ensemble spread. As such, it is hardly possible for the Monte-Carlo dropout model to produce an ensemble spread of the correct magnitude without properly scaling the resulting ensemble runs. 

However, having been trained on multiple years of atmospheric reanalysis data, if the model succeeds at all in learning the underlying physics of the atmosphere, then it is fair to assume that the model will have also learned the various possible evolutions of similar atmospheric pattern (such as possible tracks of cyclones) and thus will have learned a basic understanding of the inherent uncertainty in predicting these pattern. Thus, assessing the extend of the correlation between the predicted model uncertainty, captured in $z_{\rm std}$, as realized through Monte-Carlo dropout and the true uncertainty of the further evolution of the atmospheric parameters is a main research question to be answered.

In the present study we found experimentally that choosing $\gamma=10$ uniformly for all parameters gives reasonably well-defined probability cones that will envelop most of the test cases encountered. A more quantitative initial analysis of the results of the ensemble prediction model is presented in the following section. Still, we reserve a more careful calibration of the ensemble prediction model to future studies, where the uncertainty in the initial conditions will also be taken into account.

\subsection{Implementation}

The models were implemented in \texttt{Keras} using the \texttt{TensorFlow 2.2} backend. Training was done on a single Nvidia P100 GPU using \texttt{Google Colab}. The code is available upon request on \texttt{Github}\footnote{\url{https://github.com/abihlo/vid2vidEra5}}.

\section{Results}\label{sec:Results}

In this section we present the results obtained from training the above described \texttt{vid2vid} architecture. We first present an overall assessment of the forecast quality by computing root-mean-squared errors, the anomaly correlation coefficients and the continuous ranked probability scores. We then present a few case studies which will allow us to investigate some of the specific properties of the obtainable results using data-drive ensemble prediction systems.

\subsection{Errors}

Two standard measures for assessing the accuracy of numerical weather foreacasts of the quantity $z$ at time $t$ are the \textit{root-mean-squared error} (RMSE), defined as
\[
\textup{RMSE}_z(t) = \sqrt{\overline{(z_{\rm fcst}(t) - z_{\rm obs}(t))^2}}
\]
and the \textit{anomaly correlation coefficient} (ACC), defined as
\[
\textup{ACC}_z(t) = \frac{(\overline{z_{\rm fcst} - z_{\rm clim}})\cdot (\overline{z_{\rm obs} - z_{\rm clim}})}{\sqrt{\overline{(z_{\rm fcst}(t) - z_{\rm clim}(t))^2}}\sqrt{\overline{(z_{\rm obs}(t) - z_{\rm clim}(t))^2}} }
\]
see e.g.~\cite{wilk11a}. Here $z_{\rm fcst}$ is the forecast vector, $z_{\rm obs}$ is the observation vector (here chosen as the ERA5 reanalysis data) and $z_{\rm z_{\rm clim}}$ are the climatological values over the reanalysis time, with the bar denoting spatial averaging. 

\begin{figure}[!ht]
\centering
\begin{subfigure}{0.3\textwidth}
  \centering
  \includegraphics[width=\linewidth]{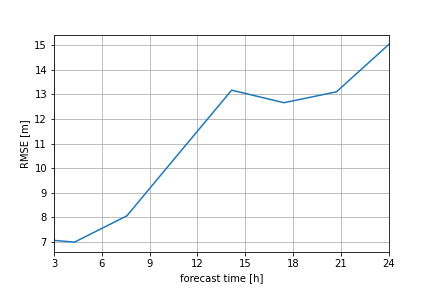}
  \caption{RMSE 500 hPa geopotential}
\end{subfigure}
\begin{subfigure}{0.3\textwidth}
  \centering
  \includegraphics[width=\linewidth]{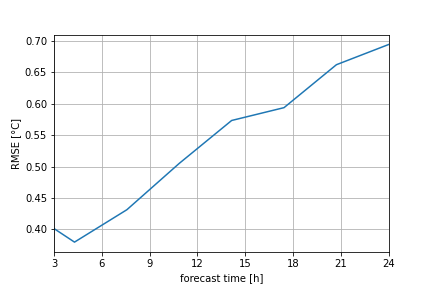}
  \caption{RMSE 2 m temperature}
\end{subfigure}
\begin{subfigure}{0.3\textwidth}
  \centering
  \includegraphics[width=\linewidth]{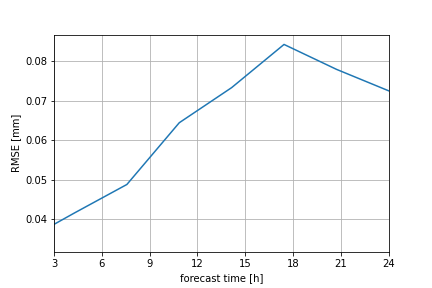}
  \caption{RMSE total precipitation}
\end{subfigure}\\
\begin{subfigure}{0.3\textwidth}
  \centering
  \includegraphics[width=\linewidth]{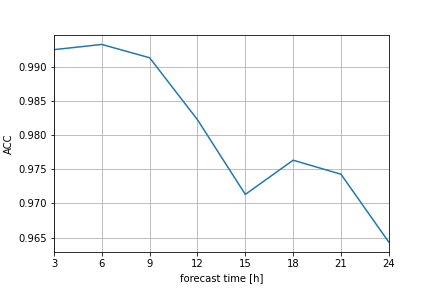}
  \caption{ACC 500 hPa geopotential}
\end{subfigure}
\begin{subfigure}{0.3\textwidth}
  \centering
  \includegraphics[width=\linewidth]{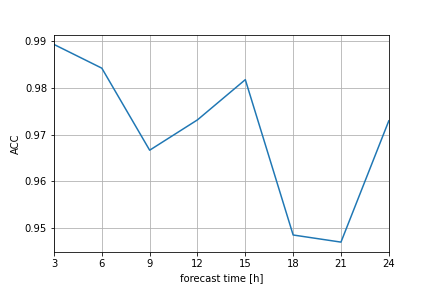}
  \caption{ACC 2 m temperature}
\end{subfigure}
\begin{subfigure}{0.3\textwidth}
  \centering
  \includegraphics[width=\linewidth]{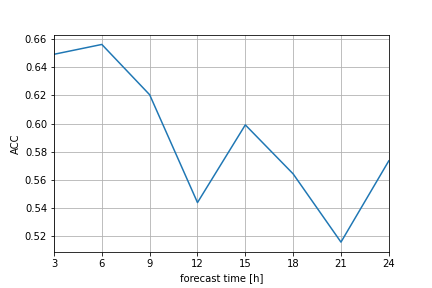}
  \caption{ACC total precipitation}
\end{subfigure}
\caption{Root-mean-squared error (RMSE) and anomaly correlation coefficient (ACC) of the \texttt{vid2vid} models evaluated for 2019.}
\label{fig:RMSEandACC}
\end{figure}

ERA5 provides monthly averaged data by the hour of the day, and we compute an approximate $z_{\rm clim}$ by averaging this data for 30 years (1989--2018). To compute the ACC for the test data we thus average the \texttt{vid2vid} results as well as the 3-hourly ERA5 reanalysis data over each month for 2019 to obtain the monthly averaged forecast and observation data by the hour of the day. The reported values in Figure~\ref{fig:RMSEandACC} are then the mean over the entire year 2019.

While the RMSE is a standard metric to measure the overall point-wise accuracy of a weather forecast, the ACC measures the ability of a forecast to predict anomalies in comparison to the climate mean. Perfect weather forecasts have an ACC of 1 while a value of 0 shows no improvement over using climatology to make a forecast. It has also been found subjectively that once the ACC is below 0.6, the resulting forecasts have no practical (synoptic) value any more. The ACC is insensitive to any bias in the forecasts. For more information see~\cite{wilk11a}.

The results for both RMSE and ACC for the three \texttt{vid2vid} networks are depicted in Figure~\ref{fig:RMSEandACC}.

The results for the geopotential are as expected the best among all three \texttt{vid2vid} results, with $\textup{ACC}_{\rm geo}$ close to 1 over the entire forecast horizon. The $\textup{ACC}_{\rm geo}$ and $\textup{RMSE}_{\rm geo}$ values for a single day prediction are competitive with standard operational numerical weather forecasts for this quantity, and in line with the results obtained in~\cite{sche19a,weyn19a}.  

The results for the two-meter temperature are almost of the same level of reliability as measured in the $\textup{ACC}_{\rm t2m}$ as those for the geopotential. In particular, $\textup{ACC}_{\rm t2m}$ does not drop far below 0.95, and interestingly does not assume its minimum value at 24 h. We have trained the associated \texttt{vid2vid} model multiple times and have received the same qualitative shape in all runs. We attribute this to the model learning to minimize the error over the entire one-shot forecasting window of 24 hours (rather than iteratively and frame-by-frame), which then seems to have found a minimum producing the particular shape of $\textup{ACC}_{\rm t2m}$. In addition, this shape may also be influenced by diurnal variations, which are obviously pronounced for the two-meter temperature and non-existent for the geopotential height. The RMSE shows a mean error of not more than $0.7^{\circ}$C which thus still gives usable weather forecasts.

The results for the total precipitation are the worst among all results. As indicated in the introduction, total precipitation is a complicated parameter that critically depends on other meteorological quantities, which have not been included in the training of the model. Another factor explaining the rather poor performance of the precipitation \texttt{vid2vid} model is that the ERA5 data, even at full available resolution, are likely not high resolution enough for the model to learn a meaningful representation of precipitation physics. Precipitation, particularly convective precipitation, in contrast to geopotential height and the two-meter temperature is usually a highly localized quantity which may require resolution of much smaller scales than those used here. As such, the low values of $\textup{ACC}_{\rm tp}$ for this parameter are expected and likely limited by the data rather than the model architecture. Indeed, this forecast has hardly any skill at all, and beyond 9 hours it is below the threshold of 0.6 marking no practical usability any more.

To get some quantitative idea of the performance of the ensemble prediction system, we use the \textit{continuous ranked probability score} (CRPS), which is a quadratic measure of the difference of the cumulative distribution function of the ensemble forecasts and the cumulative distribution function of the observation (which is a step function here), see~\cite{zamo18a} for further details. The CRPS is defined as
\[
\mathrm{CRPS} = \int_{-\infty}^\infty (P_{\rm fcst}(z)- P_{\rm obs}(z))^2\mathrm{d}z,
\]
where
\begin{align*}
&P_{\rm obs}(z) = P(z_{\rm{obs}} \leqslant z) = H(z_{\rm{obs}}-z),\\
&P_{\rm fcst}(z) = P(z_{\rm{fcst}} \leqslant z), 
\end{align*}
are the cumulative distribution functions of the observation and forecasts, respectively, with $H(z_{\rm{obs}}-z)$ being the Heavyside step function. The minimum value for the \textup{CRPS} is 0, with larger values indicating poorer performance. Note that for deterministic forecasts, the CRPS reduces to the mean absolute error.

\begin{figure}[!ht]
\centering
\begin{subfigure}{0.3\textwidth}
  \centering
  \includegraphics[width=\linewidth]{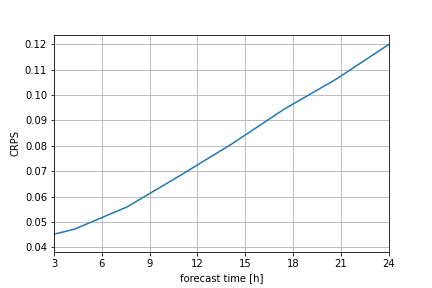}
  \caption{CRPS 500 hPa geopotential}
\end{subfigure}
\begin{subfigure}{0.3\textwidth}
  \centering
  \includegraphics[width=\linewidth]{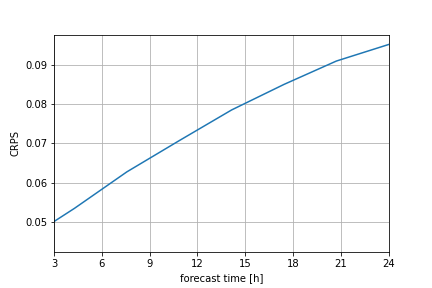}
  \caption{CRPS 2 m temperature}
\end{subfigure}
\begin{subfigure}{0.3\textwidth}
  \centering
  \includegraphics[width=\linewidth]{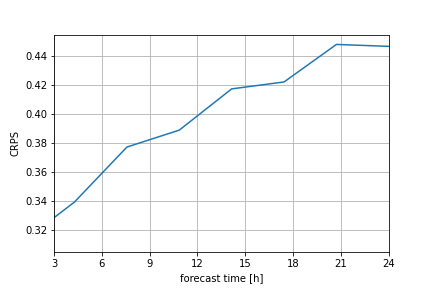}
  \caption{CRPS total precipitation}
\end{subfigure}
\caption{Continuous ranked probability score (CRPS) of the Monte-Carlo dropout \texttt{vid2vid} ensemble prediction models evaluated for 2019.}
\label{fig:CRPS}
\end{figure}

The results for the CRPS of the three \texttt{vid2vid} models using the normalized test data for 2019 are presented in Figure~\ref{fig:CRPS}. These results again quantitatively confirm that the probabilistic geopotential and temperature forecasts are the most accurate, followed by, with quite some extra margin of difference, the total precipitation forecasts. These results also yield that the \texttt{vid2vid} models ensemble prediction skills deteriorate over time, as is meteorologically meaningful, in that the uncertainty grows the longer the forecast horizon is.

\subsection{Case studies}

Here we present three case studies to showcase some individual results of the proposed ensemble prediction system. The first is post-tropical storm Lorenzo, a former Category 5 hurricane that transitioned into a post-tropical storm on October 2, 2019 moving over Ireland on October 3--4, 2019. The second is a series of extreme convective events over continental Europe on June 10--11, 2019, in the wake of storm Miguel. In meteorological terms, the first event is an advective event, the second one a convective event. Both test cases are rather challenging from a machine learning perspective, given that extreme weather events like these are by definition rare and thus not sufficiently represented in the training data. The last test case showcases the ability of the two-meter temperature prediction model to correctly capture the seasonal variations at a challenging location in the Austrian Alps.

In all ensemble plots for the various cities shown for the following case studies, \textit{ERA5} (blue) corresponds to the ground truth, \textit{ML{\_}mean} (black) corresponds to the mean of the respective \texttt{vid2vid} model and \textit{ML{\_}ens} (grey) corresponds to the scaled estimated uncertainty produced by the \texttt{vid2vid} model.

\subsubsection{Post-tropical storm Lorenzo over Ireland}

Hurricane Lorenzo was among the strongest hurricanes in the eastern or central Atlantic, see~\cite{zeli19a}. On 12 UTC on October 2, 2019 it became a post-tropical cyclone north of the Azores and brought gale force winds to the Republic of Ireland on October 4. Here we are using the \texttt{vid2vid} model to predict the geopotential height, two-meter temperature and total precipitation for the next 24 hours starting from 18 UTC October 3, 2019.

\begin{figure}[!ht]
\centering
\begin{subfigure}{\textwidth}
  \centering
  \includegraphics[width=\linewidth, trim = {0 9cm 0 9cm}, clip]{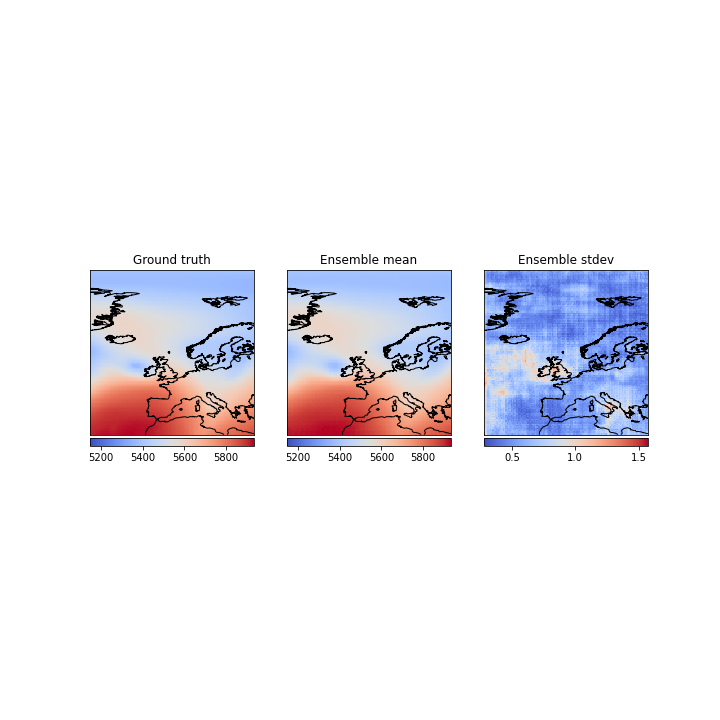}
  \caption{3 h forecast}
\end{subfigure}\\[1ex]
\begin{subfigure}{\textwidth}
  \centering
  \includegraphics[width=\linewidth, trim = {0 9cm 0 9cm}, clip]{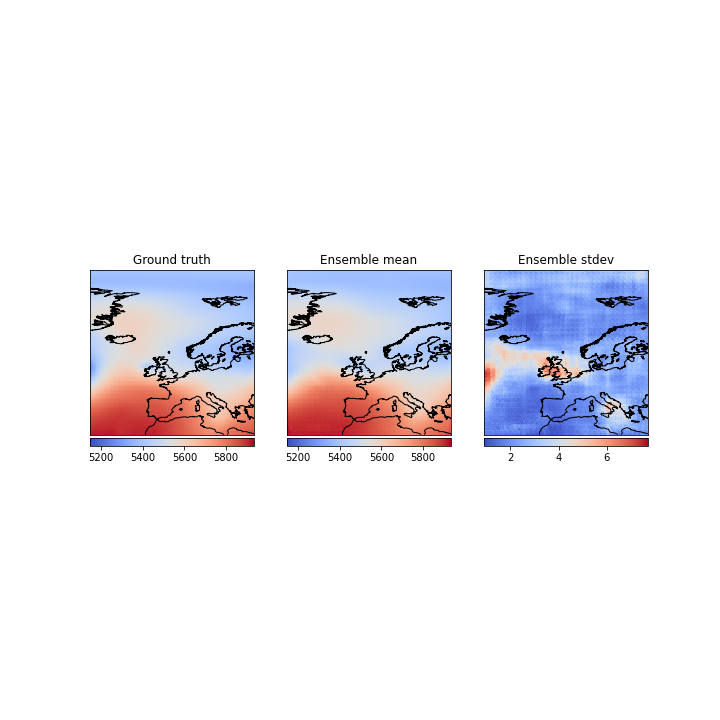}
  \caption{12 h forecast}
\end{subfigure}\\[1ex]
\begin{subfigure}{\textwidth}
  \centering
  \includegraphics[width=\linewidth, trim = {0 9cm 0 9cm}, clip]{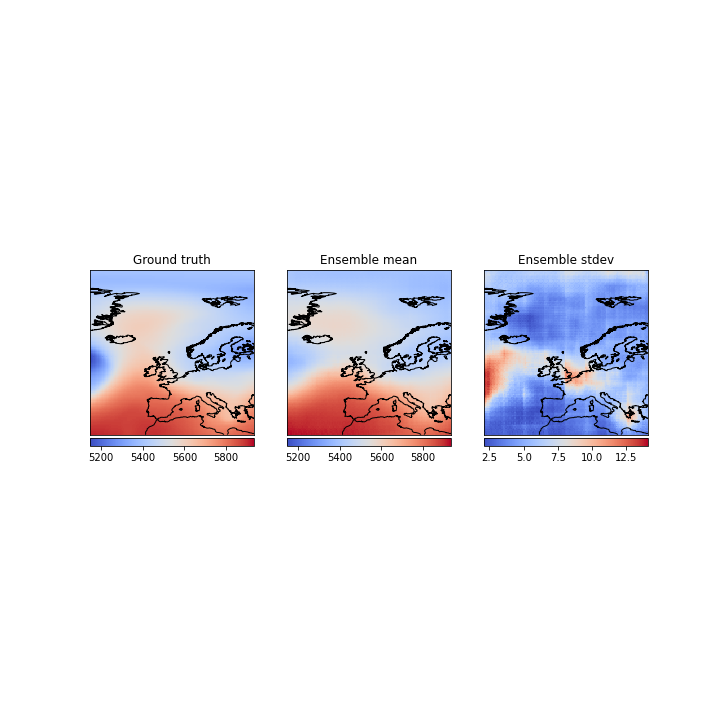}
  \caption{24 h forecast}
\end{subfigure}
\caption{Geopotential ensemble prediction for October 3--4, 2019.}
\label{fig:GeopotentialLorenzo}
\end{figure}

The evolution of the geopotential height at forecast hours 3, 12 and 24 is depicted in Figure~\ref{fig:GeopotentialLorenzo}. The core of post-tropical storm Lorenzo can be clearly seen lying west of the Irish coast at forecast time $t=3$ h, moving over Ireland and weakening at $t=12$ h, and weakening further at $t=24$ h. The \texttt{vid2vid} model can capture this evolution both qualitatively and quantitatively reasonably well, and also identifies the region over the British Isles as areas with the highest forecast uncertainty as indicated by high values of $z_{\rm std}$ at all forecast times. This is meteorologically reasonable since the path of a cyclone as well as its amplification or dissipation does exhibit an inherent degree of uncertainty.

The forecasts at $t=12$ h and $t=24$ h also highlight that the model correctly identified the Western boundary of the domain as an area of high uncertainty. This is remarkable as the model has no knowledge over the next incoming low pressure region. We interpret this result as the ability of the model to have learned the basic properties of Rossby wave dynamics, which are moving from the West to the East, possibly amplifying in the course, thus yielding a higher uncertainty in this region. Notably, no such area of uncertainty exists at the other boundaries of the domain.

\begin{figure}[!ht]
\centering
\begin{subfigure}{\textwidth}
  \centering
  \includegraphics[width=\linewidth, trim = {0 9cm 0 9cm}, clip]{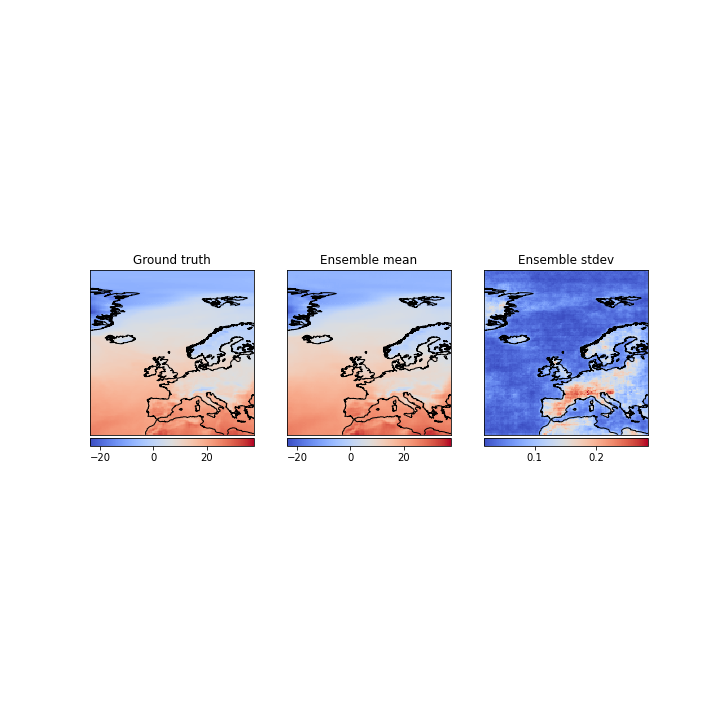}
  \caption{3 h forecast}
\end{subfigure}\\[1ex]
\begin{subfigure}{\textwidth}
  \centering
  \includegraphics[width=\linewidth, trim = {0 9cm 0 9cm}, clip]{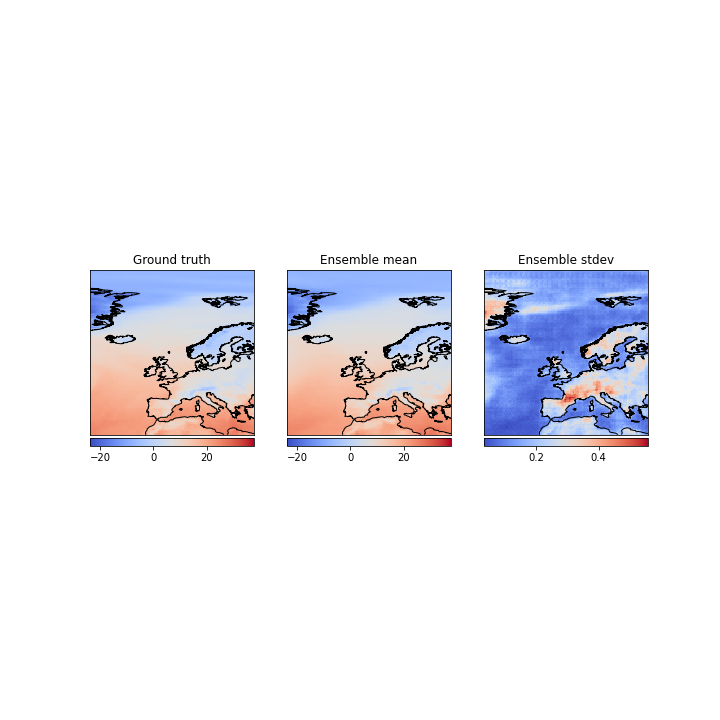}
  \caption{12 h forecast}
\end{subfigure}\\[1ex]
\begin{subfigure}{\textwidth}
  \centering
  \includegraphics[width=\linewidth, trim = {0 9cm 0 9cm}, clip]{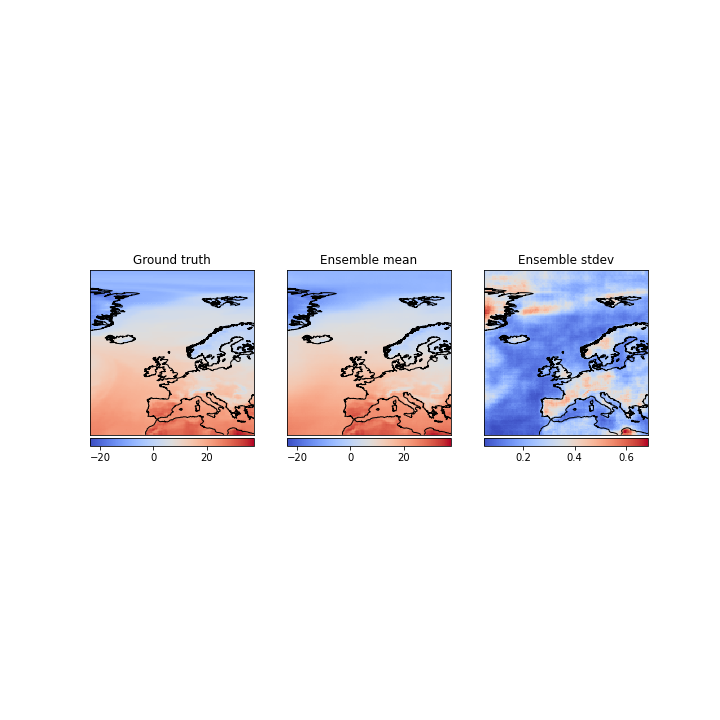}
  \caption{24 h forecast}
\end{subfigure}
\caption{Two-meter temperature ensemble prediction for October 3--4, 2019.}
\label{fig:TemperatureLorenzo}
\end{figure}

Figure~\ref{fig:TemperatureLorenzo} shows the corresponding two-meter temperature forecasts. Again the \texttt{vid2vid} model has learned to produce an overall meaningful temperature prediction at all forecast times. The highest uncertainty is generally estimated over land masses which is again physically reasonable given that the temperature over land has higher oscillations than over the ocean. At both the $t=12$ h and $t=24$ h forecasts the model again identifies the Western boundary over the North Atlantic as a region of high uncertainty, indicating that also the temperature variant of the \texttt{vid2vid} model has learned the basic physics of Rossby wave propagation and their associated temperature advections. It is also instructive to point out that the model captures that mountainous regions (such as the Alps, Iceland and Norway) are generally cooler than the surrounding regions (such as lowlands and the ocean).

\begin{figure}[!ht]
\centering
\begin{subfigure}{\textwidth}
  \centering
  \includegraphics[width=\linewidth, trim = {0 9cm 0 9cm}, clip]{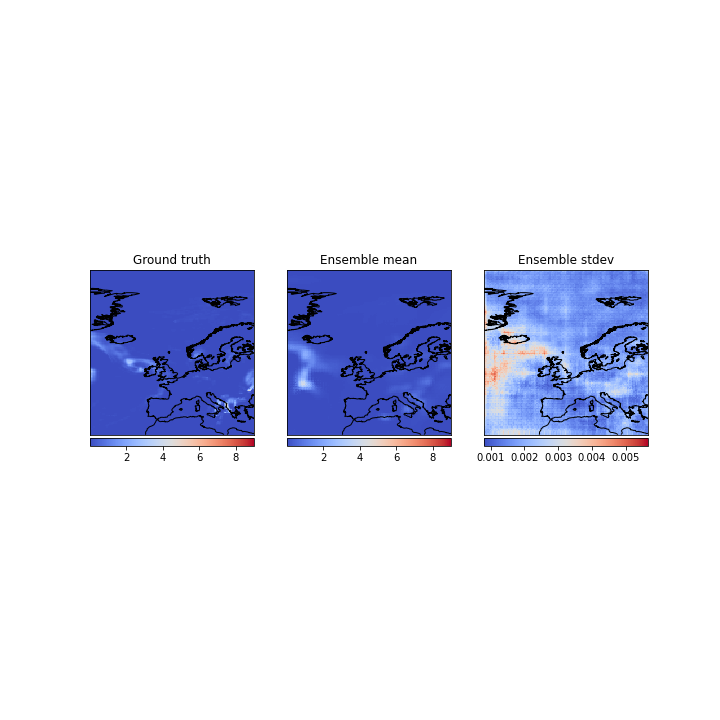}
  \caption{3 h forecast}
\end{subfigure}\\[1ex]
\begin{subfigure}{\textwidth}
  \centering
  \includegraphics[width=\linewidth, trim = {0 9cm 0 9cm}, clip]{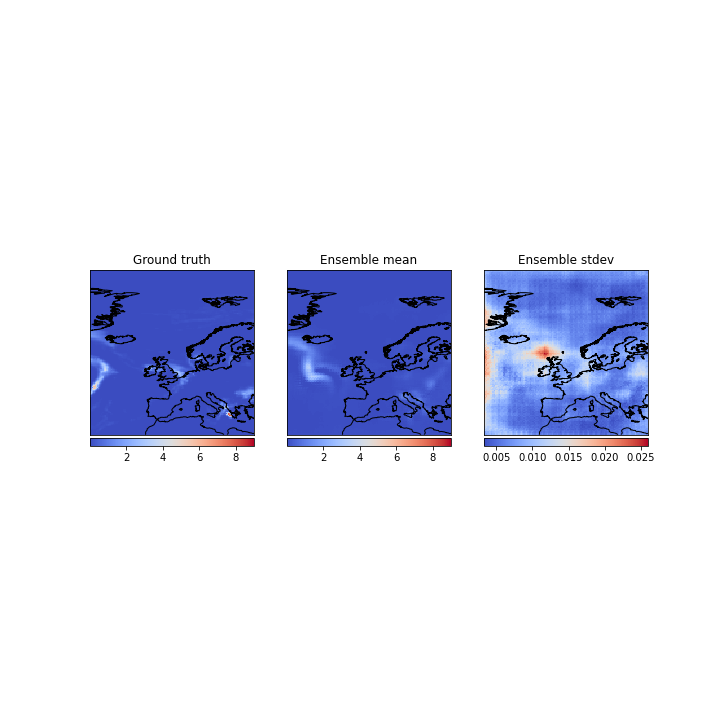}
  \caption{12 h forecast}
\end{subfigure}\\[1ex]
\begin{subfigure}{\textwidth}
  \centering
  \includegraphics[width=\linewidth, trim = {0 9cm 0 9cm}, clip]{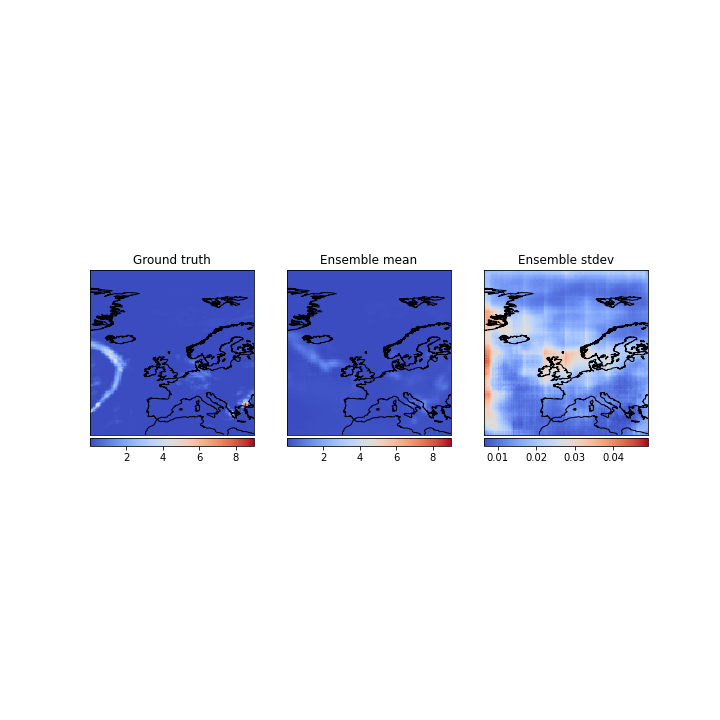}
  \caption{24 h forecast}
\end{subfigure}
\caption{Total precipitation ensemble prediction results for January 14, 2019.}
\label{fig:PrecipitationLorenzo}
\end{figure}

Figure~\ref{fig:PrecipitationLorenzo} finally shows the corresponding results for the total precipitation. As anticipated in light of the ACC results presented above, the predictions show almost no skill at all forecast times. While at forecast time $t=3$ h the model is able to identify part of the frontal precipitation system associated with post-tropical storm Lorenzo, its location is mostly incorrect. The model does produce some of the precipitation over the Adriatic Sea though. While there is little skill in the ensemble mean of the prediction, the standard deviation of the prediction correctly identifies both the area over the British Isles and the western boundary over the North Atlantic as regions of higher uncertainty. Thus, while a single forecast obtained using a data-driven forecasting model trained on precipitation data alone is unreliable at best, using it in the context of ensemble prediction may indeed improve the usability of such a model at almost no extra computational cost.

\begin{figure}[!ht]
\centering
\begin{subfigure}{.3\textwidth}
  \centering
  \includegraphics[width=\linewidth]{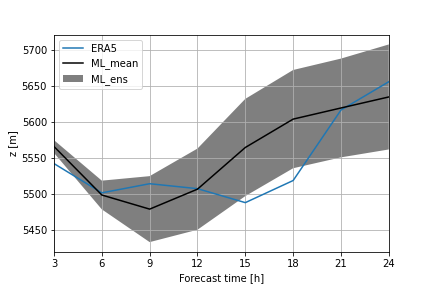}
  \caption{500 hPa geopotential}
\end{subfigure}%
\begin{subfigure}{.3\textwidth}
  \centering
  \includegraphics[width=\linewidth]{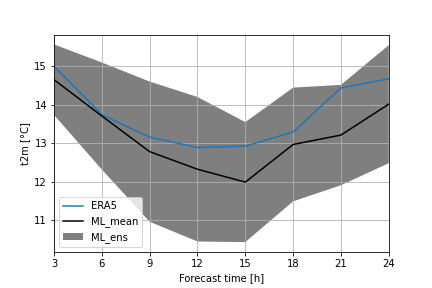}
  \caption{2 m temperature}
\end{subfigure}
\begin{subfigure}{.3\textwidth}
  \centering
  \includegraphics[width=\linewidth]{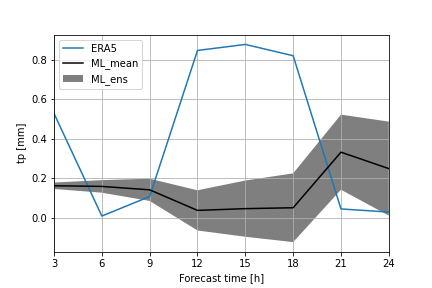}
  \caption{Total precipitation}
\end{subfigure}
\caption{Ensemble prediction for Dublin on October 3--4, 2019.}
\label{fig:DublinLorenzo}
\end{figure}

\begin{figure}
\centering
\begin{subfigure}{.3\textwidth}
  \centering
  \includegraphics[width=\linewidth]{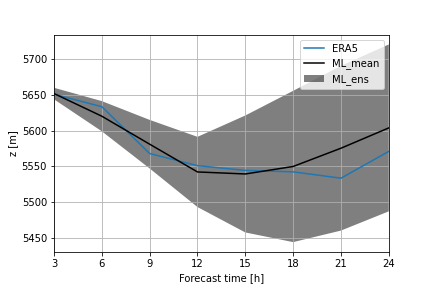}
  \caption{500 hPa geopotential height}
\end{subfigure}%
\begin{subfigure}{.3\textwidth}
  \centering
  \includegraphics[width=\linewidth]{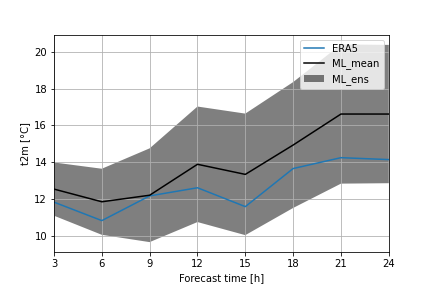}
  \caption{2 m temperature}
\end{subfigure}
\begin{subfigure}{.3\textwidth}
  \centering
  \includegraphics[width=\linewidth]{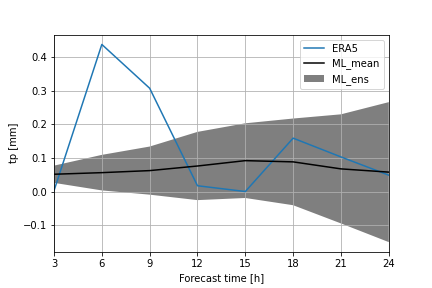}
  \caption{Total precipitation}
\end{subfigure}
\caption{Ensemble prediction for London on October 3--4, 2019.}
\label{fig:LondonLorenzo}
\end{figure}

In Figures~\ref{fig:DublinLorenzo} and~\ref{fig:LondonLorenzo} we show the time series of the forecast parameters at Dublin and London, respectively. Both time series show an overall correct trend in geopotential height and two-meter temperature but fail to capture the evolution of the precipitation at these two locations. It is also instructive to note that all ensembles tend to estimate higher uncertainties at later forecast times, which is typically meaningful.

\subsubsection{Severe thunderstorms over continental Europe}

As our next case study we investigate a time frame of unsettled weather over continental Europe from June 10 to June 11, 2019. Storm Miguel has hit Portugal, Spain and France and a region of unsettled weather downstream of this storm produced hail and thunderstorms over Germany, Poland, Northern Italy and Croatia as described by~\cite{puci19}. We aim to forecast the weather on June 11, 2019.

\begin{figure}[!ht]
\centering
\begin{subfigure}{\textwidth}
  \centering
  \includegraphics[width=\linewidth, trim = {0 9cm 0 9cm}, clip]{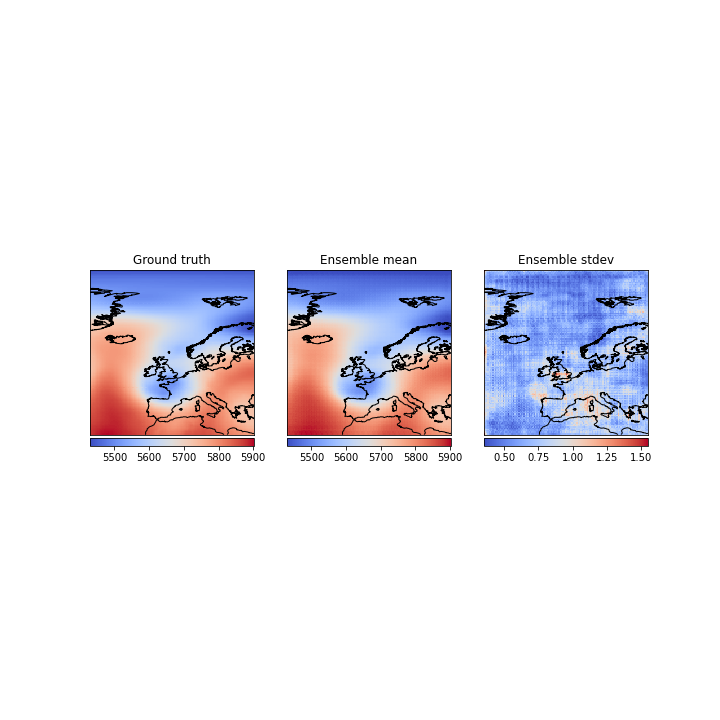}
  \caption{3 h forecast}
\end{subfigure}\\[1ex]
\begin{subfigure}{\textwidth}
  \centering
  \includegraphics[width=\linewidth, trim = {0 9cm 0 9cm}, clip]{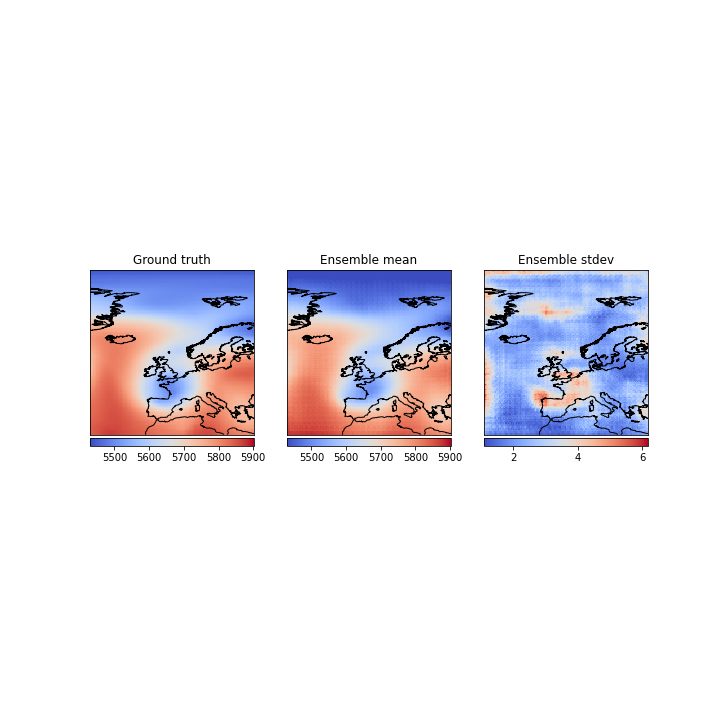}
  \caption{12 h forecast}
\end{subfigure}\\[1ex]
\begin{subfigure}{\textwidth}
  \centering
  \includegraphics[width=\linewidth, trim = {0 9cm 0 9cm}, clip]{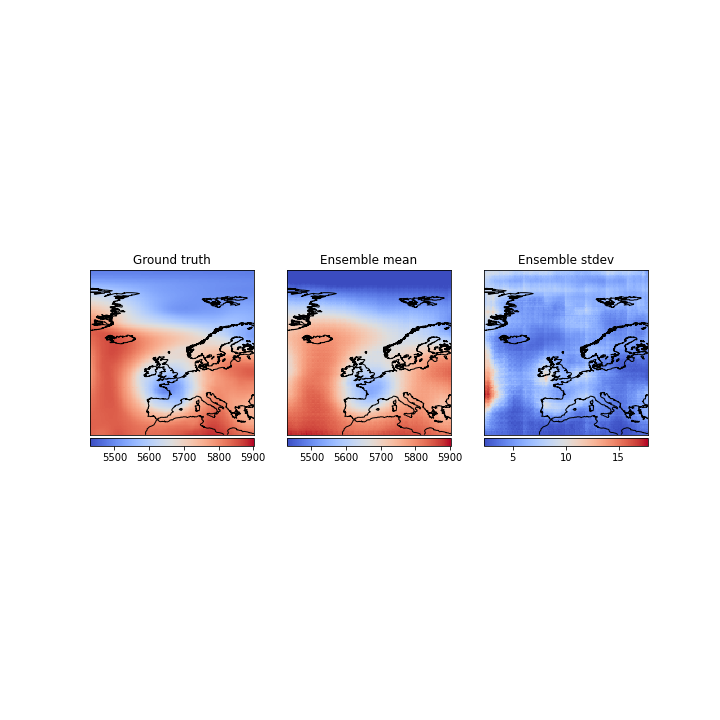}
  \caption{24 h forecast}
\end{subfigure}
\caption{Geopotential ensemble prediction for June 11, 2019.}
\label{fig:GeopotentialMiguel}
\end{figure}

Figure~\ref{fig:GeopotentialMiguel} shows the geopotential height forecast produced by the $\texttt{vid2vid}$ ensemble prediction model. As in the previous case, the model shows an excellent agreement with the true evolution of the 500 hPa geopotential height surface on this day. The model captures the slow evolution of the core of storm Miguel from off the coast of France well and identifies the regions around this storm as areas of higher uncertainty at all forecast times $t=3$ h, $t=12$ h and $t=24$ h. The model also again captures the uncertainty over the western boundary over the North Atlantic.

\begin{figure}[!ht]
\centering
\begin{subfigure}{\textwidth}
  \centering
  \includegraphics[width=\linewidth, trim = {0 9cm 0 9cm}, clip]{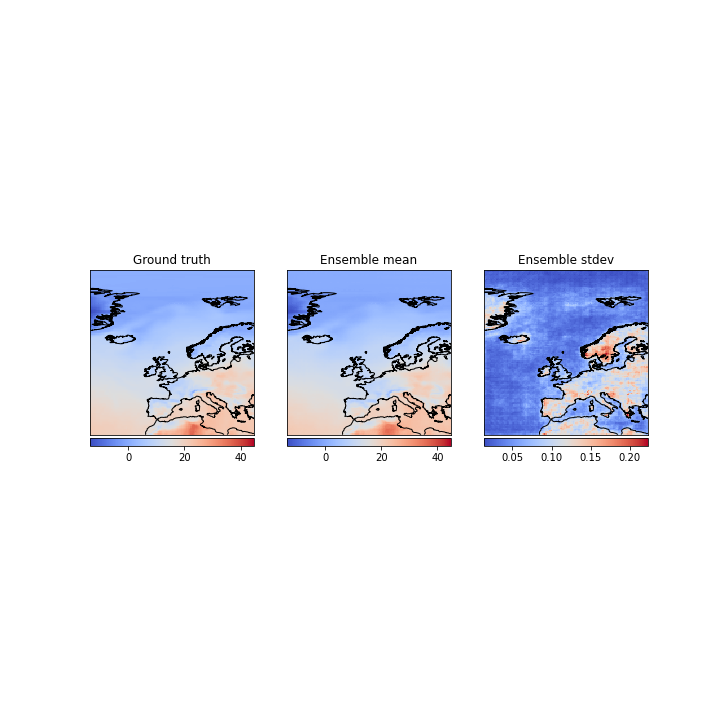}
  \caption{3 h forecast}
\end{subfigure}\\[1ex]
\begin{subfigure}{\textwidth}
  \centering
  \includegraphics[width=\linewidth, trim = {0 9cm 0 9cm}, clip]{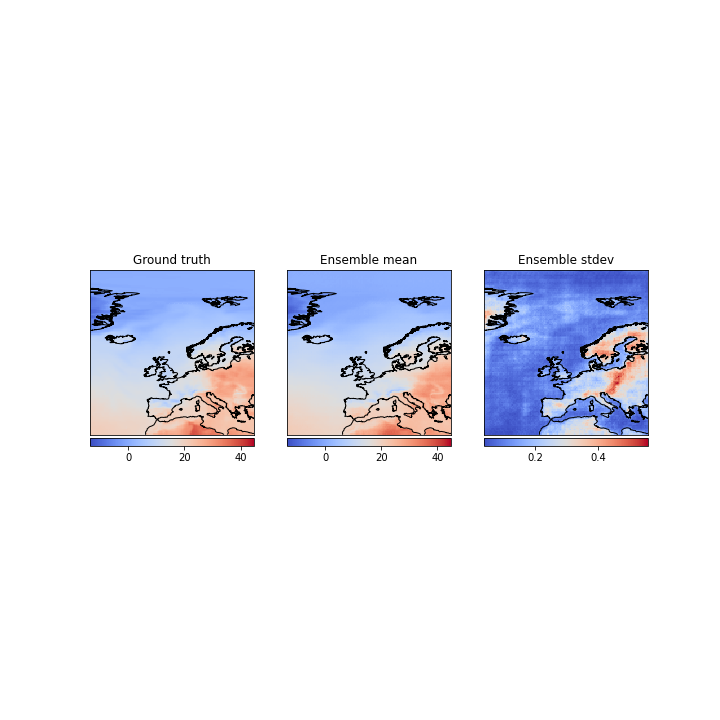}
  \caption{12 h forecast}
\end{subfigure}\\[1ex]
\begin{subfigure}{\textwidth}
  \centering
  \includegraphics[width=\linewidth, trim = {0 9cm 0 9cm}, clip]{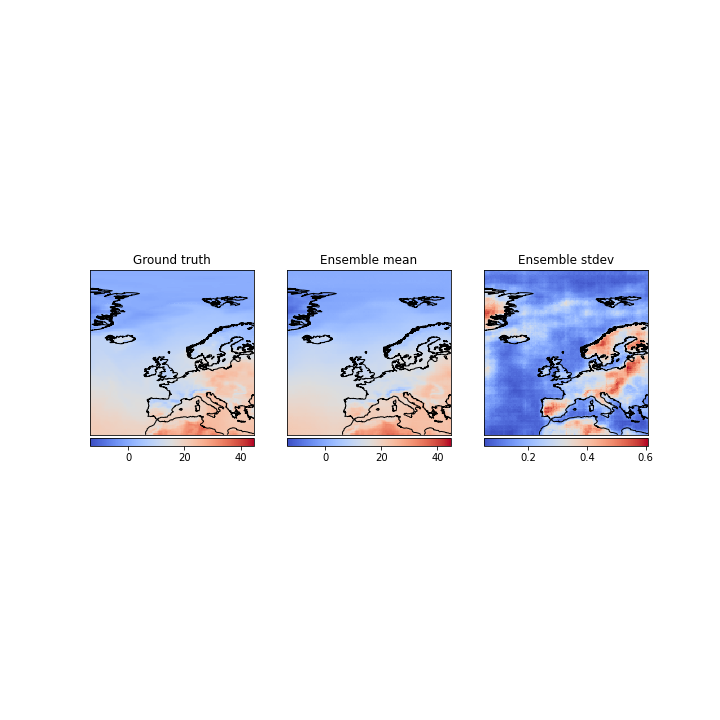}
  \caption{24 h forecast}
\end{subfigure}
\caption{Two-meter temperature ensemble prediction for June 11, 2019.}
\label{fig:TemperatureMiguel}
\end{figure}

Figure~\ref{fig:TemperatureMiguel} shows the associated two-meter temperature predictions. Visually these forecasts agree reasonably well, with the most interesting feature being high values of uncertainty forecast over the Baltic states, Poland, Hungary, Slovenia, Croatia and Northern Italy, as well as over Portugal and Spain. Most of these regions experienced significant weather this day.

\begin{figure}[!ht]
\centering
\begin{subfigure}{\textwidth}
  \centering
  \includegraphics[width=\linewidth, trim = {0 9cm 0 9cm}, clip]{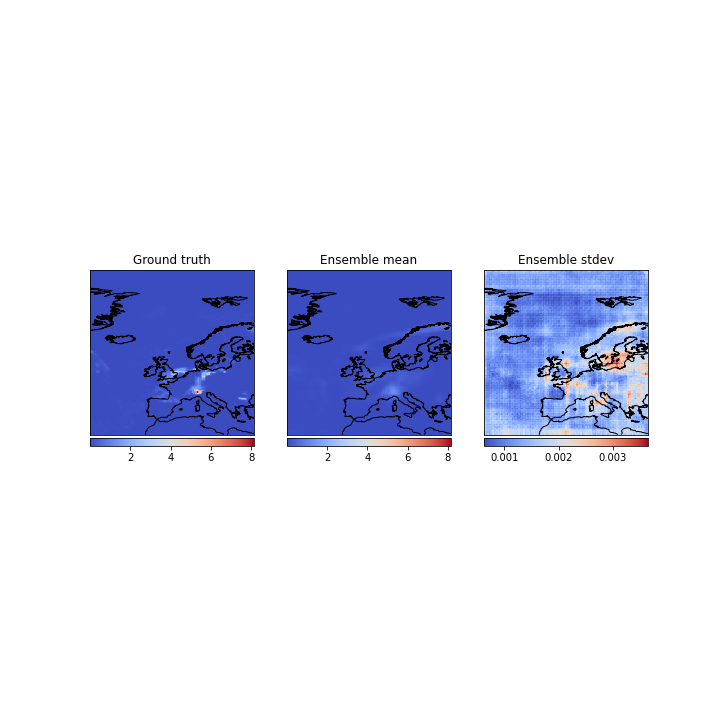}
  \caption{3 h forecast}
\end{subfigure}\\[1ex]
\begin{subfigure}{\textwidth}
  \centering
  \includegraphics[width=\linewidth, trim = {0 9cm 0 9cm}, clip]{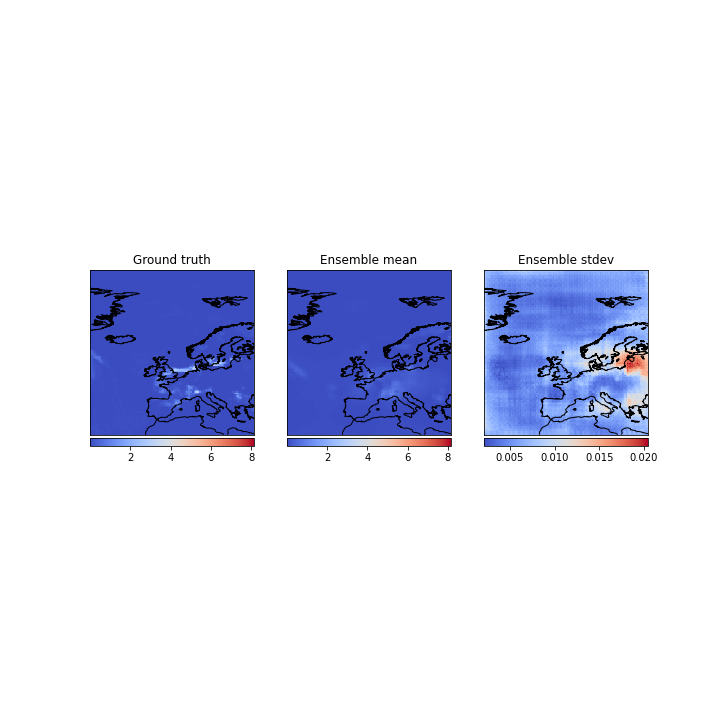}
  \caption{12 h forecast}
\end{subfigure}\\[1ex]
\begin{subfigure}{\textwidth}
  \centering
  \includegraphics[width=\linewidth, trim = {0 9cm 0 9cm}, clip]{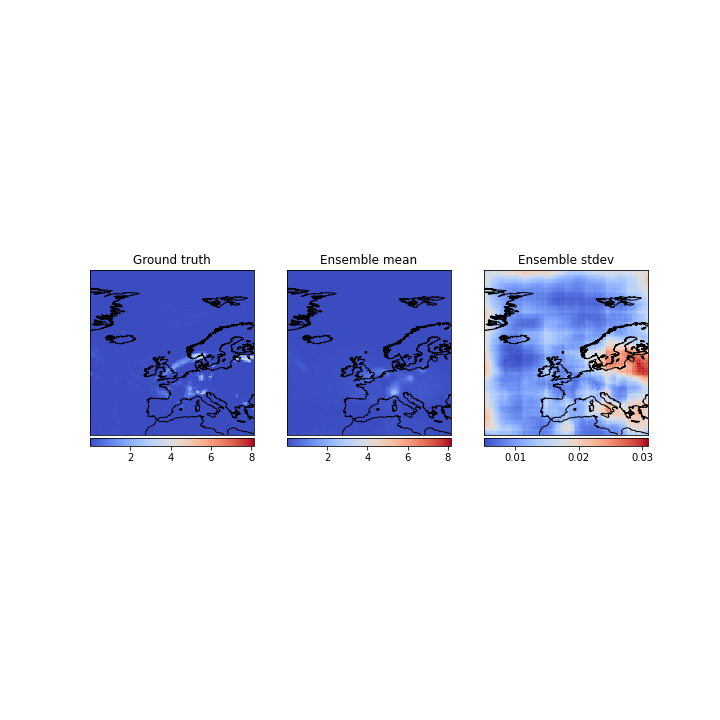}
  \caption{24 h forecast}
\end{subfigure}
\caption{Total precipitation ensemble prediction for June 11, 2019.}
\label{fig:PrecipitationMiguel}
\end{figure}

Figure~\ref{fig:PrecipitationMiguel} shows the corresponding total precipitation forecast. Here the results are slightly better than in the previous case study, in that the model correctly identifies the precipitation region in Northern Italy at $t=3$ h, $t=12$ h and $t=24$ h, albeit the quantities being significantly underestimated. Interestingly, the model also identifies the region over the Baltic Sea as an area of uncertainty in the prediction. Overall, these results are again illustrating that there is little skill in precipitation forecasts trained on precipitation alone, but with the ensemble model providing some extra useful information not captured in a single forecast alone. 

\begin{figure}[!ht]
\centering
\begin{subfigure}{.3\textwidth}
  \centering
  \includegraphics[width=\linewidth]{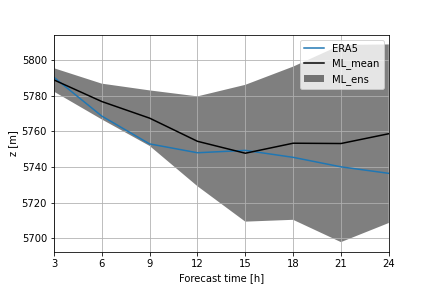}
  \caption{500 hPa geopotential height}
\end{subfigure}%
\begin{subfigure}{.3\textwidth}
  \centering
  \includegraphics[width=\linewidth]{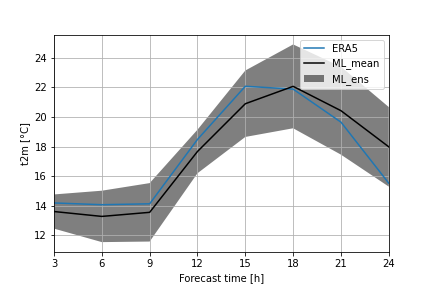}
  \caption{2 m temperature}
\end{subfigure}
\begin{subfigure}{.3\textwidth}
  \centering
  \includegraphics[width=\linewidth]{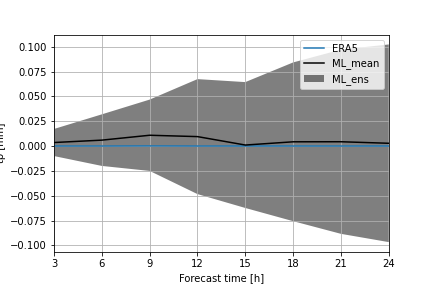}
  \caption{Total precipitation}
\end{subfigure}
\caption{Ensemble prediction results for Lisbon on June 11, 2019.}
\label{fig:LisbonMiguel}
\end{figure}

\begin{figure}[!ht]
\centering
\begin{subfigure}{.3\textwidth}
  \centering
  \includegraphics[width=\linewidth]{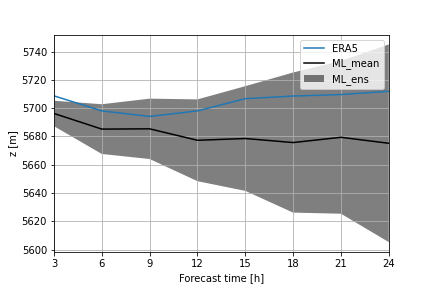}
  \caption{500 hPa geopotential height}
\end{subfigure}%
\begin{subfigure}{.3\textwidth}
  \centering
  \includegraphics[width=\linewidth]{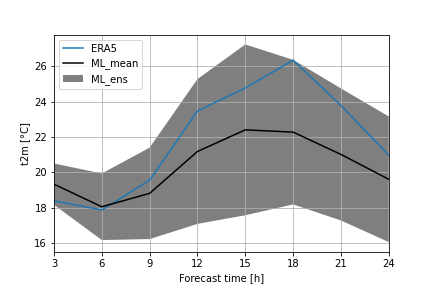}
  \caption{2 m temperature}
\end{subfigure}
\begin{subfigure}{.3\textwidth}
  \centering
  \includegraphics[width=\linewidth]{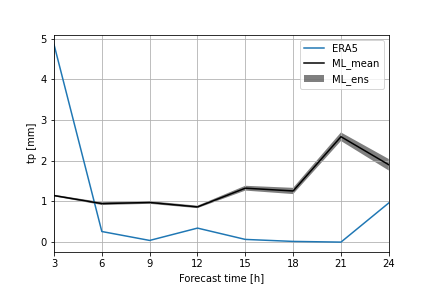}
  \caption{Total precipitation}
\end{subfigure}
\caption{Ensemble prediction results for Milan on June 11, 2019.}
\label{fig:MilanMiguel}
\end{figure}

In Figures~\ref{fig:LisbonMiguel} and~\ref{fig:MilanMiguel} we show the time series of predictions for Lisbon and Milan on this day. As before, the geopotential height and temperature predictions are in excellent agreement with the true values on that day, with the precipitation results being mostly unusable. From these individual ensemble results it is particularly interesting to point out that the model has higher confidence in the prediction at Lisbon than it has at Milan, as is indicated by a quite substantially smaller ensemble spread at Lisbon than at Milan. This may be interpreted as the model having learned that there are less uncertainties to be expected in an advective regime (such as Lisbon experience on that day) than in a convective regime (such as Milan experience on that day). This makes sense meteorologically since predicting the advection of meteorological quantities is generally easier than predicting the life-cycle of a convective event.

\subsection{Case study of seasonal variation of two-meter temperature}

As one last case study we investigate the ability of the \texttt{vid2vid} model to predict the time series of temperature throughout different seasons. We pick the town of Innsbruck in Austria as a location here, which due to its location in a deep Alpine valley exhibits challenging orographic variations in its temperature. In the winter months Innsbruck often experiences inversion events, where the temperature on the surface (underneath a cloud layer) is lower than the temperature at the surrounding mountains (above the cloud layer). In the summer time, due to reduced volume in the valley, temperature can get higher than in the surrounding lowlands. Since there is no such systematic variation in the geopotential height, we do not investigate the geopotential height forecasts here. While the ERA5 data are likely not high-resolution enough to capture all these local effects accurately, as is visually clear from the previous two cases studies they do include some noticable orographic variations. Assessing whether the temperature \texttt{vid2vid} model has learned these orographic variations and their seasonal variations is thus still a meaningful benchmark.

\begin{figure}[!ht]
\centering
\begin{subfigure}{0.3\textwidth}
  \centering
  \includegraphics[width=\linewidth]{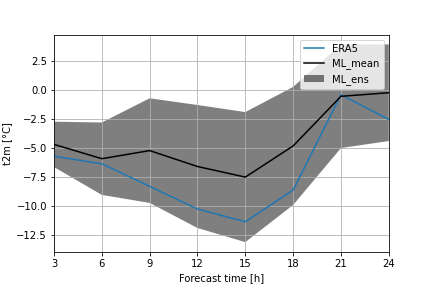}
  \caption{January 14--15, 2019}
\end{subfigure}
\begin{subfigure}{0.3\textwidth}
  \centering
  \includegraphics[width=\linewidth]{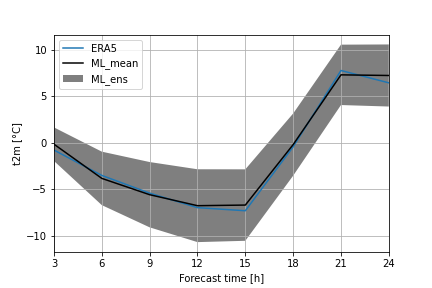}
  \caption{February 14--15, 2019}
\end{subfigure}
\begin{subfigure}{0.3\textwidth}
  \centering
  \includegraphics[width=\linewidth]{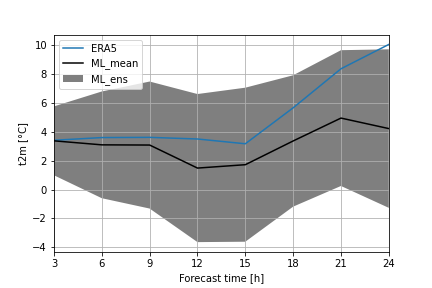}
  \caption{March 14--15, 2019}
\end{subfigure}
\begin{subfigure}{0.3\textwidth}
  \centering
  \includegraphics[width=\linewidth]{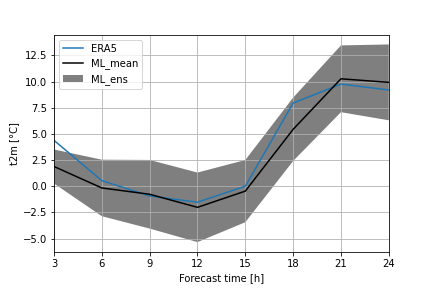}
  \caption{April 14--15, 2019}
\end{subfigure}
\begin{subfigure}{0.3\textwidth}
  \centering
  \includegraphics[width=\linewidth]{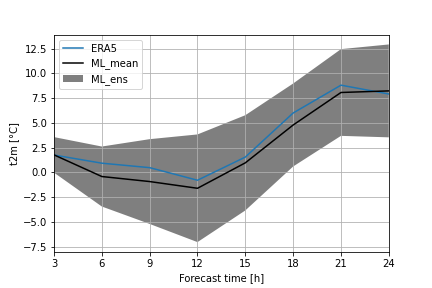}
  \caption{May 14--15, 2019}
\end{subfigure}
\begin{subfigure}{0.3\textwidth}
  \centering
  \includegraphics[width=\linewidth]{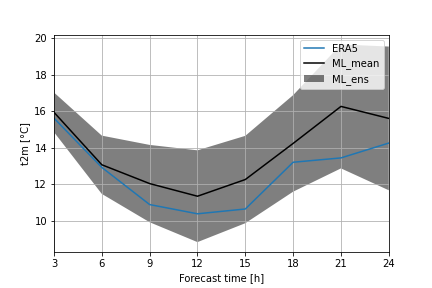}
  \caption{June 14--15, 2019}
\end{subfigure}
\begin{subfigure}{0.3\textwidth}
  \centering
  \includegraphics[width=\linewidth]{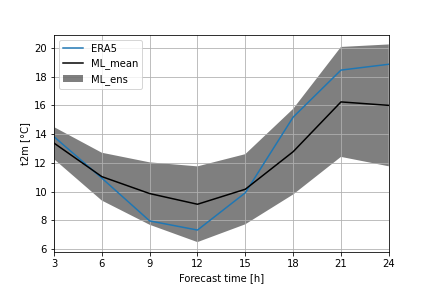}
  \caption{July 14--15, 2019}
\end{subfigure}
\begin{subfigure}{0.3\textwidth}
  \centering
  \includegraphics[width=\linewidth]{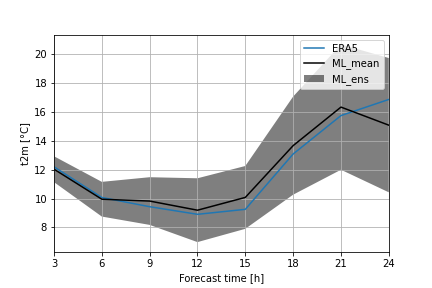}
  \caption{August 14--15, 2019}
\end{subfigure}
\begin{subfigure}{0.3\textwidth}
  \centering
  \includegraphics[width=\linewidth]{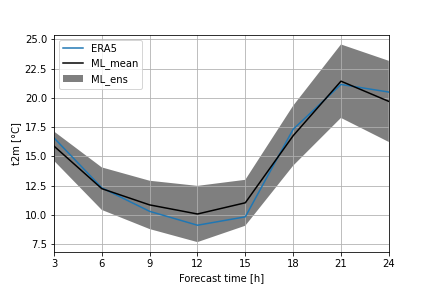}
  \caption{September 14--15, 2019}
\end{subfigure}
\begin{subfigure}{0.3\textwidth}
  \centering
  \includegraphics[width=\linewidth]{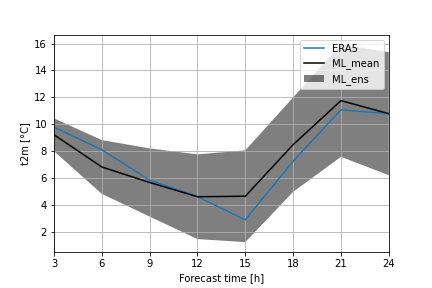}
  \caption{October 14--15, 2019}
\end{subfigure}
\begin{subfigure}{0.3\textwidth}
  \centering
  \includegraphics[width=\linewidth]{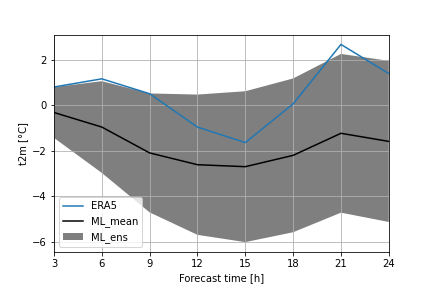}
  \caption{November 14--15, 2019}
\end{subfigure}
\begin{subfigure}{0.3\textwidth}
  \centering
  \includegraphics[width=\linewidth]{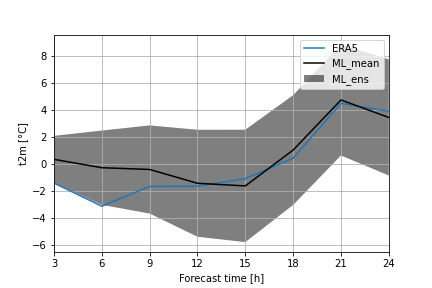}
  \caption{December 14--15, 2019}
\end{subfigure}
\caption{Two-meter temperature prediction for the 15th of each month in 2019 for Innsbruck.}
\label{fig:T2mIBK}
\end{figure}

Figure~\ref{fig:T2mIBK} shows the ensemble forecasts for Innsbruck at the 15th of each month for 2019. The model has clearly learned to predict temperature at this challenging location and shows high skill in each month. Again it is interesting to point out that the estimated uncertainty is different for each day, with the ensemble mean typically closely following the observed temperature at the same location.

\section{Conclusions}\label{sec:Conclusions}

In this paper we have investigated the use of a conditional generative adversarial network in learning the conditional probability distribution underlying the physics of the atmospheric system. We have chosen the height of the 500 hPa pressure level, the two-meter temperature and the total precipitation as test parameters, and trained three separate versions of the exact same \texttt{vid2vid} architecture using 4 years of ERA5 reanalysis data to predict the weather over the next 24 hours for the period of 1 year.

The obtained results for all parameters as assessed through the ACC, the RMSE, the CRPS and individual case studies, show the potential ability of the architecture to learn the basic physical processes as well as some of the inherent uncertainty underlying the evolution of the atmospheric system. In particular the addition of Monte-Carlo dropout improves the reliability of the results by providing a cheaply available error estimate for the machine learning based weather forecasts, which produced meaningful meteorological results. 

Immediate room for improvement of the obtained results can be anticipated by including further features in the training of these networks. In particular precipitation is closely correlated with regions of unstable air masses and frontal zones, and including features capturing the vertical stratification of the atmosphere may help in improving the results for this parameter. Using global (or at least hemispherical) data rather than just a limited spatial domain as in our study would alleviate the ill-posedness of the upstream boundary condition, which implies a higher uncertainty at later forecasting times in the vicinity of the boundary regions, which would have to be dealt with, should longer range forecasts be investigated. Still, developing a limited area data-driven ensemble prediction model that correctly identified the upstream boundary condition as area of higher uncertainty allows us to conclude that the model has learned some of the physical properties such as the propagation of Rossby waves from the data.

Regarding the rather disappointing albeit expected results for the precipitation forecasting \texttt{vid2vid} model we should also like to again point out that the ERA5 data are likely not high-resolution enough to allow the model to learn meaningful signals from this data. Precipitation is mostly localized, and in particular convective precipitation takes place on scales far smaller than could be represented in the ERA5 data. Re-training a precipitation \texttt{vid2vid} on higher resolution data alone might already significantly improve the quality of the associated forecasting system. Coupling high-resolution data with additional features linked to precipitation will then likely further improve the obtainable results.

While the issue of further improvement of the forecast results could likely be dealt with using more features and data, the question as to whether standard Monte-Carlo dropout is the most effective strategy for obtaining data-driven ensemble weather predictions should be investigated further in more detail. This concerns in particular the relation between the standard deviation of an ensemble of data-driven weather forecasts based on Monte-Carlo dropout alone and the predictability of the state of the atmosphere over the intended forecast horizon. For the time being, the results obtained at least demonstrate that Monte-Carlo dropout, if properly tuned, may be a feasible and cheap initial approach in quantifying the uncertainty in data driven-weather forecasts and providing some error estimates for single model runs. The case studies we presented indicate that the ensemble prediction models have learned to associate uncertainty in the prediction with dominant weather pattern such as cyclones as well as that they have correctly identified the predominant westerly flow on the Northern hemisphere. 

Still, to further assess the ability of machine learning methods to produce meaningfully calibrated meteorological ensemble predictions it will be necessary to include an estimate of the uncertainty of the current state of the atmosphere into training of the employed neural network architectures, which we did not do in the present paper. We aim to tackle the problem of adding this uncertainty information of the initial state of the atmosphere to the models in the near future.

\section*{Acknowledgements}

The author is grateful to Jason Brownlee and the developers of \texttt{TensorFlow} for providing codes for the \texttt{pix2pix} models adapted in this study. This research was undertaken thanks to funding from the Canada Research Chairs program, the InnovateNL LeverageR{\&}D program and the NSERC Discovery Grant program.

{\footnotesize\setlength{\itemsep}{0ex}

}

\end{document}